\documentclass[a4paper]{article}
\usepackage{latexsym}
\begin{document}
\title{Some Remarks on Real-Time Turing Machines\\
(Preliminary Report)}
\author{Holger Petersen\\
Reinsburgstr.~75\\
70197 Stuttgart
}

\maketitle

\begin{abstract}
The power of real-time Turing machines using sublinear space is investigated.
In contrast to a claim appearing in the literature, such machines can accept
non-regular languages, even if working in deterministic mode. While maintaining
a standard binary counter appears to be impossible in real-time, we present a
guess and check approach that yields a binary representation of the input length.
Based on this technique, we show that unary encodings of languages accepted in
exponential time can be recognized by nondeterministic real-time Turing machines.
\end{abstract}

\newcommand{\qed}{$\Box$}
\newtheorem{lemma}{Lemma}
\newtheorem{theorem}{Theorem}
\newtheorem{corollary}{Corollary}
\newtheorem{observation}{Observation}
\newtheorem{claim}{Claim}

\section{Introduction}
From one of the earliest papers on computational complexity \cite{Stearns65}, it is known
that there are lower space bounds for accepting non-regular languages. In the case of 
deterministic two-way machines space proportional to $\log\log n$ is required,
while for one-way machines the bound is $\log n$. An interesting claim of a stronger gap
for real-time machines appears in \cite{Bruda10}, namely that even nondeterministic machines
of this kind require linear space for accepting non-regular sets.  
In contrast to this claim we show here that  deterministic real-time Turing machines can 
accept non-regular languages in sublinear space. While it seems impossible to count in a
standard way with a real-time machine, we develop a counting technique for 
nondeterministic real-time Turing machines such that
they have access to a binary representation of the input length before having consumed 
more than half of their input. We use this technique for a general result about
single-letter languages acceptable by nondeterministic real-time Turing machines.
A consequence of this result is that the primes in unary can be accepted by 
nondeterministic real-time Turing machines disproving a conjecture appearing in the
literature.

\section{Discussion of Bruda's Model}
The definition of Turing machines given in \cite{Bruda10}
deviates in several ways from the definition of classical papers 
like \cite{Book70}. There is a single halt state $h$ (in
comparison to a set of such states in \cite{Book70}) and
no transition from this state is possible. More significant is 
the fact that the definition of acceptance in \cite{Bruda10} 
is independent of the contents of input- and work-tapes. 
A consequence of this aspect of its formal definition is 
that a Turing machine working in real-time
accepts any extension of a shorter accepted input, 
since an accepting configuration is reached on the longer
input string as well.  In particular, every
single-letter language accepted by machines of this kind is 
either empty or co-finite and
thus regular. While this does not contradict Bruda's claim that 
languages accepted by real-time Turing machines in sublinear space
are regular, we will now present a language accepted 
in real-time and $O(\sqrt{n})$ space.
Let the set of states of the real-time Turing machine $M_s$ 
with one input-tape and one work-tape be 
$K = \{ q_0, q_1, q_2, q_3 \}$ with $q_0$ the initial state. 
The alphabet is $\Sigma = \{ \#, a, 0, 1 \}$ (where \# is the blank)
and the transition function mapping $K \times \Sigma^2$ to
$(K \cup \{ h\}) \times \{ R, L, N\}^2 \times \Sigma^2$ is given by 
\begin{eqnarray*}
\delta(q_0, a, \#) & = & (q_1, R, R, a, 0)\\
\delta(q_0, a, 0) & = & (q_0, R, R, a, 0)\\
\delta(q_1, a, \#) & = & (q_2, R, L, a, 0)\\
\delta(q_1, b, \#) & = & (h, R, L, b, 0)\\
\delta(q_2, a, 0) & = & (q_2, R, L, a, 0)\\
\delta(q_2, a, \#) & = & (q_3, R, L, a, 0)\\
\delta(q_3, a, \#) & = & (q_0, R, R, a, 0)\\
\delta(q_3, b, \#) & = & (h, R, L, b, 0)
\end{eqnarray*}
The recognition of perfect squares is based on the well-known sum of
the first $k$ odd numbers:
$$\sum_{i=0}^{k - 1} 2i+1 = k^2$$
Machine $M_s$ marks successively segments of the tape having odd lengths.
For a prefix $a^nb$ with $n = k^2$ it will sweep over a segment with $2k-1$ 
cells while reading $a^n$ and mark an additional cell for the $b$. For
$n$ strictly between perfect squares no additional spce is required 
during the sweep. Therefore, the space usage of $M_s$ is bounded by 
$2\sqrt{n}$ and thus sublinear. Next we argue that    
$$S = \{ a^nbx \mid \mbox{ $n$ is a perfect square, $x \in \{ a, b \}^*$}\}$$
is not regular.
The right quotient $S/b\{ a, b \}^* = 
\{ a^n \mid \mbox{ $n$ is a perfect square}\}$ is a well-known
non-regular language (Exercise 4.1.2(a) in \cite{Hopcroft79}).
This quotient would be regular if $S$ was, as follows from the closure of the
regular languages under quotient (Theorem~9.13 from \cite{Hopcroft69}).

\section{A Nondeterminsitic Counting Technique}
In the present section we adopt the standard definition of acceptance by final state and empty 
storage. 
\begin{lemma}\label{counter}
There is a nondeterministic real-time Turing machine $M_c$ (counter) with the following properties:
\begin{enumerate}
\item $M_c$ writes down a guess of $n$ in binary on a designated work tape before having read $n/2$
 symbols of an input of length $n$.
\item $M_c$ enters a special state at the end of its computation if and only if the guessed value was
correct.
\item $M_c$ uses $O(\log n)$ space in every computation.
\end{enumerate}
\end{lemma}
Proof. Machine $M_c$ executes several processes in parallel making use of multiple tapes and 
the cross product of states of Turing machines implementing these processes. 
The work tapes will be referenced by the following names:
\begin{itemize}
\item CURRENT: Tape contents of a single-tape Turing machine $M_d$ defined below
 representing the number of input symbols processed by $M_c$.
\item FINAL: Guess of the configuration $M_d$ reaches when $M_c$ has read its entire input.
\item WORK: Copy of the guessed configuration. 
\item LENGTH: Stores input length computed from guessed configuration on WORK.
\item DIFF: Unary counter keeping track of the number of differences between CURRENT and FINAL.
\end{itemize}
One process using tape CURRENT is the
simulation of the deterministic single-tape machine $M_d$ (no input tape) that maintains
a binary counter:
\begin{eqnarray*}
\delta(q_0, \#) & = & (q_1, L, \#)\\
\delta(q_0, 0) & = & (q_0, R, 0)\\
\delta(q_0, 1) & = & (q_0, R, 1)\\
\delta(q_1, \#) & = & (q_0, R, 1)\\
\delta(q_1, 0) & = & (q_0, R, 1)\\
\delta(q_1, 1) & = & (q_1, L, 0)
\end{eqnarray*}
State $q_1$ propagates a carry and $q_0$ moves the head back to the least 
significant digit.
The simulation is carried out on tape CURRENT by $M_c$.
Notice that $M_d$ has no halt state, since we are only interested
in its configuration when the input of $M_c$ has been read completely.
This configuration includes the head position and the internal state of 
$M_d$ and we extend the alphabet $\{ \#, 0, 1\}$ of $M_d$ with symbols 
$\{ \#, 0, 1\} \times \{ 0, 1\}$, where the second component represents 
state and head position of $M_d$. Whenever $M_d$ writes a digit 
on a new tape cell (except for the two most significant ones),
$M_c$ guesses a symbol $\alpha$ from the extended alphabet of $M_d$
and writes $\alpha$ onto tapes FINAL and WORK, on which the tape heads
move in parallel with $M_d$'s head. The unary counter DIFF is increased
when $\alpha$ differs from the corresponding symbol on CURRENT. 
On tapes CURRENT, FINAL and WORK identical head movements are carried out
and each time the scanned symbols become equal or different by the 
simuklation of $M_d$, the unary counter DIFF is adjusted accordingly.

If the two most significant
digits are touched by $M_d$ (which it has to guess), 
a new phase of processing starts. While 
FINAL is treated in the same way as in the first phase, $M_d$ remembers 
a guess of the most significant digit it its finite control and 
writes this digit onto WORK (it cannot be written onto FINAL because
$M_c$ necessarily moves its head to the right after the carry has been propagated
and FINAL still executes identical head movements). Next $M_c$ continues
the simulation of $M_d$ starting from the configuration on WORK until
its head is located on the blank to the right of the least significant digit
on WORK. The number of steps executed is counted in binary on tape LENGTH.
Now the number on LENGTH is subtracted from the number on WORK and
all proper prefixes of the binary number on WORK are added to the number on
tape LENGTH. Finally this number is doubled (which simply means a concatenation of 0).

We claim that LENGTH stores the binary encoding of
the number of steps that $M_d$ executes until reaching
the configuration stored on FINAL. For configurations with $M_d$'s head 
to the right of the least significant digit this is easily checked, since a 
prefix increases by one for every carry propagation. This involves 
two crossings of the right border of the prefix. We subtracted 
the number of steps until such a configuration is reached, which 
adjusts the count.

In parallel the simulatioon of $M_d$ continues and $M_c$ enters the
designated state if DIFF stores 0. 

The configuration stored on CURRENT includes a binary counter that increases
at most once per input symbol. This shows an $O(\log n)$ space bound for
CURRENT and the other tapes that depend on CURRENT. \qed

For a binary word $w \in 1\{ 0, 1\}^*$ define a padded string $\mbox{pad}(w)$ as follows:
\begin{eqnarray*}
\mbox{pad}(1) & = & a \\
\mbox{pad}(w'0) & = & \mbox{pad}(w')^2  \\
\mbox{pad}(w'1) & = & \mbox{pad}(w')^2a 
\end{eqnarray*}

We generalize padding to a language $L$ by 
$\mbox{pad}(L) = \{ \mbox{pad}(w) \mid w \in L \}$.

\begin{theorem}
Let $L\subseteq 1\{ 0, 1\}^*$. If $L \in \mbox{NTIME}(2^n)$ then $\mbox{pad}(L)$ is
accepted by a nondeterministic real-time Turing machine.
\end{theorem}
Proof. By assumption there is a nondeterministic Turing machine $M$
accepting $L$ in time $2^n$. Given an input of the form $a^k$, a simulator 
first uses the technique from Lemma~\ref{counter} for guessing $k$ in binary.
Then its starts a simulation of $M$ on this binary string with a speed up in
order to compensate the time needed for guessing $k$. If $M$ accepts and 
the guess was correct, the input $a^k$ is accepted. \qed

We can conclude from the previuos theorem that the conjecture from \cite{Book70}
that 
$$\{ a^p \mid \mbox{ $p$ is prime}\}$$
cannot be accepted in real-time is wrong. By Pratt's result
\cite{Pratt75} the primes in binary notation are in NP and therefore
clearly accepted in time $2^n$.

\end{document}